# Digital Signal Processing Using Deep Neural Networks

Brian Shevitski[1, ⸸, *], Yijing Watkins[1,⸸], Nicole Man[1] and Michael Girard[1, *]
[1]Pacific Northwest National Laboratory, Richland, WA 99352 United States
[*]To whom correspondence should be addressed: brian.shevitski@gmail.com, michael.girard@pnnl.gov
[⸸]these authors contributed equally to this work

Abstract
   *Currently there is great interest in the utility of deep neural networks (DNNs) for the physical layer of radio frequency (RF) communications. In this manuscript, we describe a custom DNN specially designed to solve problems in the RF domain. Our model leverages the mechanisms of feature extraction and attention through the combination of an autoencoder convolutional network with a transformer network, to accomplish several important communications network and digital signals processing (DSP) tasks. We also present a new open dataset and physical data augmentation model that enables training of DNNs that can perform automatic modulation classification, infer and correct transmission channel effects, and directly demodulate baseband RF signals.*

Engineering the fifth generation (5G) of mobile communications networks is uniquely challenging due to the complexity of the typical modern wireless radio frequency (RF) environment. Due to the ubiquity of mobile and wireless devices, a typical 5G system needs to handle a huge volume of complex, heterogeneous data, presenting new obstacles in the allocation and management of network resources[1].

Currently, there is great interest in the feasibility of embedding machine learning (ML) directly into a communications network to combat issues that arise in a crowded and diverse RF environment. Deep neural networks (DNNs) are currently the dominant ML architecture and have revolutionized ML model performance in the last decade across many domains including computer vision (CV)[2–7], natural language processing (NLP)[8–11], and content recommendation[12–14].

DNNs have been applied with varying degrees of success to several tasks in the physical layer of the RF communications domain[15,16]. Previous studies have primarily focused on automatic modulation classification using a variety of different DNN architectures[17]. Common ML models used for image classification in the CV domain have also proven effective for modulation classification in the RF domain, such as standard convolutional neural networks (CNNs)[18] and residual neural networks (ResNets)[17]. Limited research has been conducted on the efficacy of model architectures that are more specialized to the RF domain by using long short-term memory networks (LSTMs) for modulation type classification[19,20]. The LSTM architecture is "time-aware" and has proven invaluable in the analysis of time-series data across a number of domains[21–23].

Previous studies have shown the viability of using DNNs for digital RF signal processing. Signal demodulation using several different ML architectures has been demonstrated under a limited range of conditions[24–29]. Completely learned DNN end-to-end communications systems have been demonstrated[30,31], but these systems can be difficult to train and may require complex protocol schemes to enable assured communication links. In general, research combining ML with RF is still nascent, both in assessing what tasks ML can accomplish or improve upon and in developing tools aligned with the RF modality.

In this manuscript we outline a novel deep learning architecture, custom-designed for various RF communications network tasks. We show that DNNs can be used to directly demodulate digital baseband signals with high accuracy. We also demonstrate the first use of deep learning networks for several digital signal processing tasks, including inference of key signal properties and transmission channel and parameters.

Supervised deep learning generally requires copious amounts of accurately labeled data in order to train models that perform at a high level. While an open dataset of RF signals does exist[32], a majority of the effects that we are seeking to quantify are either not present or not labeled in the most commonly used open source RF data set. With this in mind, we designed our own dataset and physical transmission channel model in PyTorch for use in training models.

Our custom dataset is generated using the framework outlined in Fig. 1(a) which schematically shows how data moves through a generic RF communications system. First, a stream of random bits is generated at a virtual transmitter (Tx), and a non-return-to-zero (NRZ), time-domain encoding is created. The NRZ encoding is mapped onto one of 13 distinct constellations to create a complex baseband digital modulation signal. The baseband signal is then passed through a root-raised cosine (RRC) filter for pulse shaping. This final, filtered, baseband signal at the virtual transmitter is denoted Tx and is kept for future steps in the analysis pipeline.

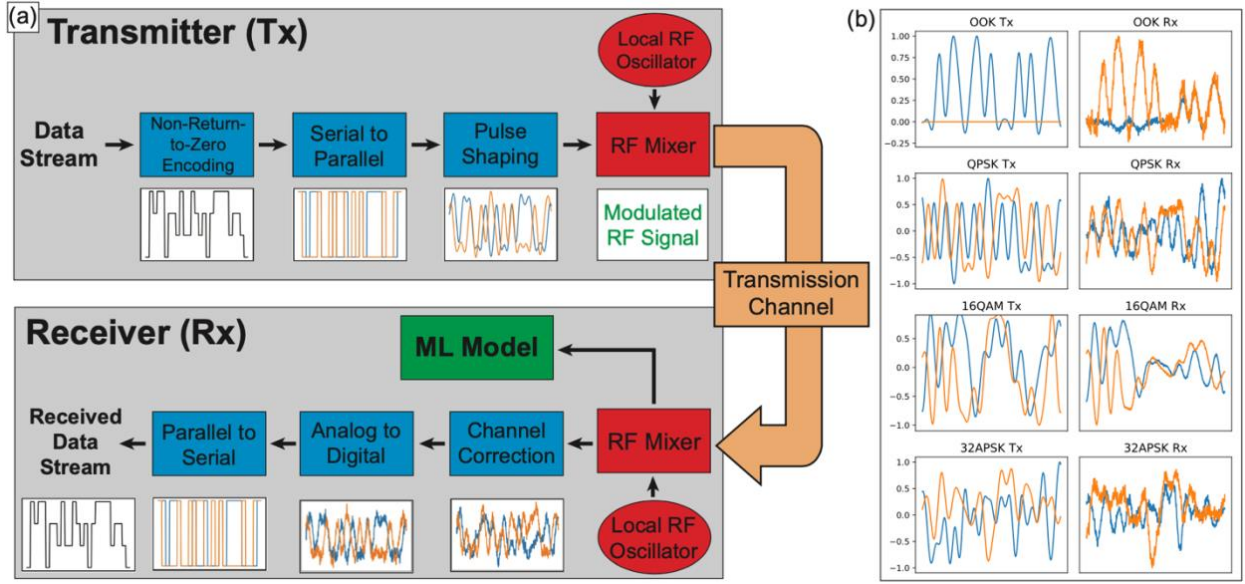

**Fig. 1** A generic RF communications system, consisting of a transmitter (Tx), a receiver (Rx), and a transmission channel, with a simplified signal processing pipeline and example data is shown in (a). A data stream is used to generate a complex baseband signal at Tx which is used to modulate a radio frequency carrier wave. The modulated carrier is then sent to Rx though the transmission channel. A copy of the Tx baseband signal, perturbed by the transmission channel, is recovered from the modulated carrier received by Rx. After further processing, a copy of the data stream is recovered at Rx. Our machine learning (ML) models sit at Rx and take the channel perturbed baseband signal as input, as indicated by the green rectangle in (a). Examples of complex baseband signals before (left) and after (right) propagation through a simulated transmission channel are shown in (b) for several different digital modulation schemes.

In order to properly simulate a realistic RF environment, the Tx signal must propagate through a simulated transmission channel. The term "transmission channel" generally encompasses any effects that cause the signal at the receiver, denoted Rx, to be different from Tx. Our physical channel model includes the most important effects: local RF oscillator phase and frequency offset, additive white gaussian noise (AWGN), and Rayleigh fading (see Methods for more details). The choice of transmission channel parameters can be tuned to simulate a transmission channel that ranges from mild to harsh depending on how robust a model must be against the RF environment.

Fig. 1(b) shows examples of Tx and Rx data for four different modulation types. Each data example is generated as needed and is accompanied by a suite of important labels and metadata such as message bits, modulation type, transmission channel parameters, number of time samples per message symbol, etc. The custom design of the dataset and transmission channel allows us to perform a wide variety of communications network and digital signals processing tasks using supervised deep learning models.

Because of their resounding success in the CV domain, a handful of DNN architectures (VGG, ResNet, etc.) are often immediately co-opted when building a ML model in a new domain. We propose a novel model, designed specifically for various tasks in the RF domain, referred to throughout as the "hybrid model" (Fig. 2). The hybrid model aims to improve performance for RF applications in two ways. First, the model learns a latent representation which is simultaneously used both for denoising/reconstructing the input baseband RF data and for some digital signal processing (DSP) task. This shared representation leads to solutions which inform one another and an improvement in performance. Second, we make use of a transformer network, which is particularly well-suited to solving problems in the time-domain[8,9,33,34].

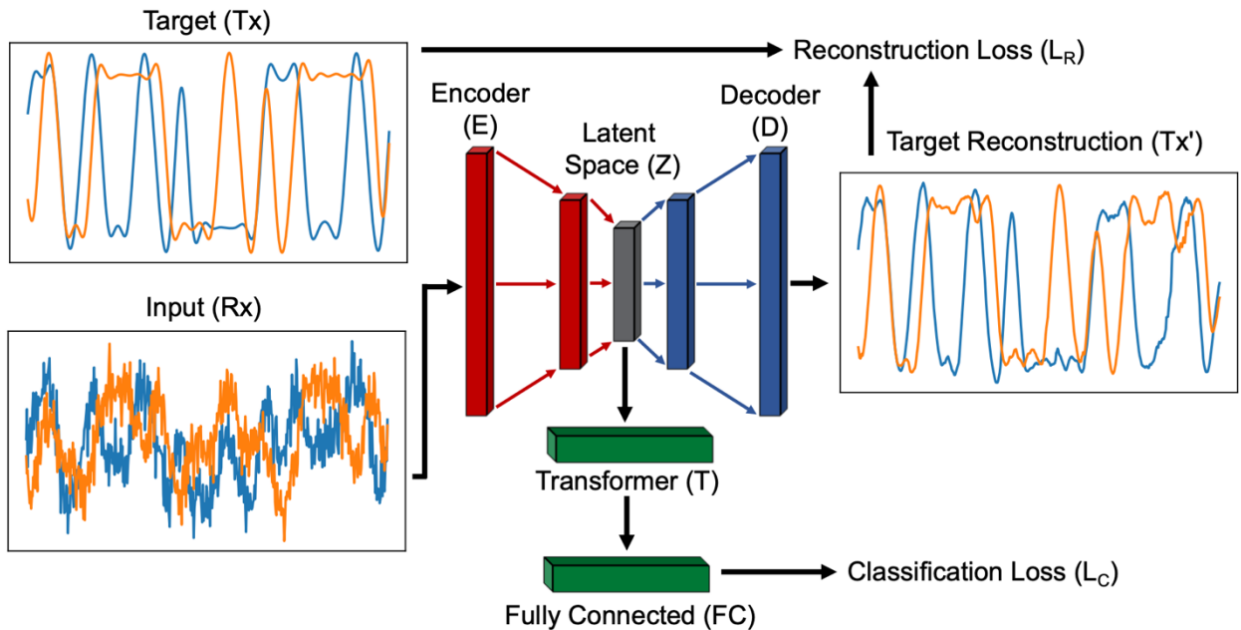

**Fig. 2** Schematic of the hybrid autoencoder/transformer model. The encoder takes a raw digital baseband signal, perturbed by a transmission channel, from a receiver (Rx) as input and compresses it into a latent representation, z. The decoder transforms the z back to the original shape of the input. The output of the decoder (Tx') is used to reconstruct the unperturbed baseband signal originally sent by the transmitter (Tx) by minimizing the loss between Tx and Tx'. The latent vector is used as input for a transformer network followed by fully connected layers. The final output of the fully connected classifier layers (C) is trained by minimizing the loss between the classifier output and known labels for the data. Once trained, the output of the classifier is used to infer various properties about the underlying data and the transmission channel.

The hybrid model architecture is composed of an encoder/decoder pair (E/D) and a classifier network (C). The encoder network consists of convolutional neural network (CNN) layers which map an input baseband RF signal (Rx) onto a latent space representation (Z). The decoder network (D), another series of CNN layers, transforms Z into a reconstructed baseband RF signal. The output of D (Tx') is a denoised and reconstructed RF signal. The latent representation is also used as input for C, the output of which is the predicted label of Rx for the network task at hand.

Our classifier network uses a transformer (we use the reformer implementation[35] of the transformer architecture) whose output is flattened and fed into fully connected neural network layers. The transformer, which has enabled great performance increases in the domain of NLP[8,9], performs an all-to-all comparison of elements which implies a utility in the analysis of sequences or other data with long-range correlation, e.g. time series data[33,34]. To our knowledge, our proposed hybrid model is the first to combine an auto-encoder convolutional network with a transformer network to attempt to solve various tasks in the RF domain.

Our custom synthetic dataset, coupled with a physical transmission channel model, allows us to use Rx baseband data as input for a DNN model and to use a wide variety of important signal and channel parameters, as well as the original Tx signal itself, as targets or labels (model outputs). Fig. 2 schematically outlines how the hybrid model is trained to perform the various RF communications network tasks presented throughout this manuscript. The Rx signal is used as input for the autoencoder network (a reasonable assumption for how a real-world ML communications system would function) and the Tx signal is used as the target for the decoder reconstruction loss function. The classifier network loss function uses the various signal and channel parameters as targets, depending on the task the model is being trained to perform. For more details on model training and loss functions, see Methods.

Previous studies have been primarily focused on the task of automatic modulation classification[17,19,20,36]. As an initial proof of concept of the hybrid models capabilities, we train the model to predict the modulation type label of our simulated baseband RF signals under a particularly harsh transmission channel. For this experiment, we allow the phase offset to vary between 0-360 degrees, the frequency offset to vary up to 1% of the data rate, and the dimensionless Rayleigh fading parameter, η, to vary between 0.1-1.0 (see Methods), with all parameters randomly generated for each data example. The results of this experiment are shown in Fig. 3. Because of our custom end-to-end data generation and channel simulation, we can easily measure the performance of the model for a given signal-to-noise ratio. Figs. 3(a) and 3(b) show confusion matrices at energy per symbol to noise power spectral density ratios of $Es/N0 = 0$ and 10 dB, respectively. The curve in Fig. 3(c) shows the mean model accuracy, averaged across all modulation types at each noise level. For more details on data generation, transmission channel simulation, and model training, see Methods.

Next, we turn our attention to a more novel DSP task using the hybrid model, directly inferring transmission channel parameters. To perform this task, we limit the input data strictly to messages of the quadrature phase-shift keying (QPSK) modulation type (a common digital modulation scheme used in many applications, including WiFi and Bluetooth®). We use a slightly less harsh (denoted medium) transmission channel than for the modulation classification task. Because of the four-fold rotational symmetry of the QPSK constellation, phase shifts of more than ±45 degrees are unresolvable from observations of the baseband signal alone. For this reason, we eliminate Rayleigh fading from the channel entirely (fading introduces a random phase shift to the baseband signal) and limit the carrier phase offsets to the range of ±45 degrees.

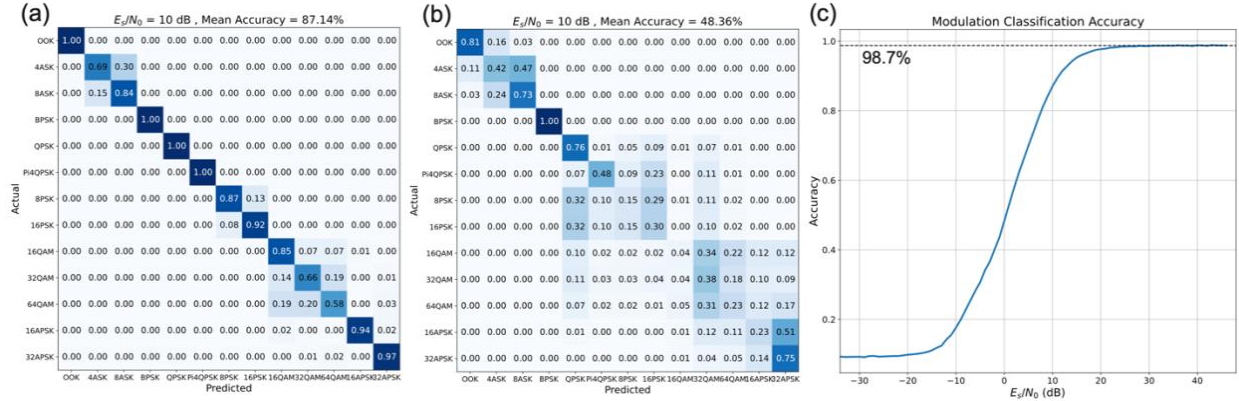

**Fig. 3** Confusion matrices showing performance of the hybrid model at automatic modulation type classification for Es/N0 of 10 dB (a) and 0 dB (b). The curve in (c) shows the mean accuracy of the hybrid model across all classes for Es/N0 values between –30 to 40 dB. For high signal to noise ratios, the hybrid model has a mean accuracy of 98.7% when identifying the modulation type of an unknown signal, as shown in (c).

Results of the channel parameter regression are shown in Fig. 4. The hybrid model can simultaneously and explicitly regress the local oscillator phase offset (Fig. 4(a)), local oscillator frequency offset (Fig. 4(b)), and signal-to-noise level (Fig. 4(c)) for a wide range of these parameters. To our knowledge, this is the first demonstration of the use of a DNN to directly infer transmission channel properties. More experimental details, including information on model dimensions, initialization, and training, can be found in Methods.

The results in Fig. 4 clearly show that the hybrid model can explicitly regress channel parameters. These parameters can then be used to reconstruct the original transmitted signal from the data at Rx using either a physical model or traditional analog electronics and signal processing methods. One key feature of the hybrid model is the encoder/decoder pair and the ability to denoise and reconstruct the original Tx signal. This implies that the model can implicitly perform the previously mentioned two-step process (learn channel parameters from Rx signal, then use them to transform Rx signal into Tx signal) in a single step (directly learn original Tx signal). Fig. 5 shows example Rx, Tx, and reconstruction signals from the regression model discussed previously that illustrate this property. Figs. 5(a) shows the Rx (green lines), Tx (blue/orange lines), and Tx reconstruction (dashed black lines) for the in-phase and quadrature components, respectively, of an example QPSK signal. Fig. 5(c) shows the error between Tx and Tx reconstruction for both components, scaled by the RMS amplitude of Tx, with a mean error of less than 8 percent (in these units).

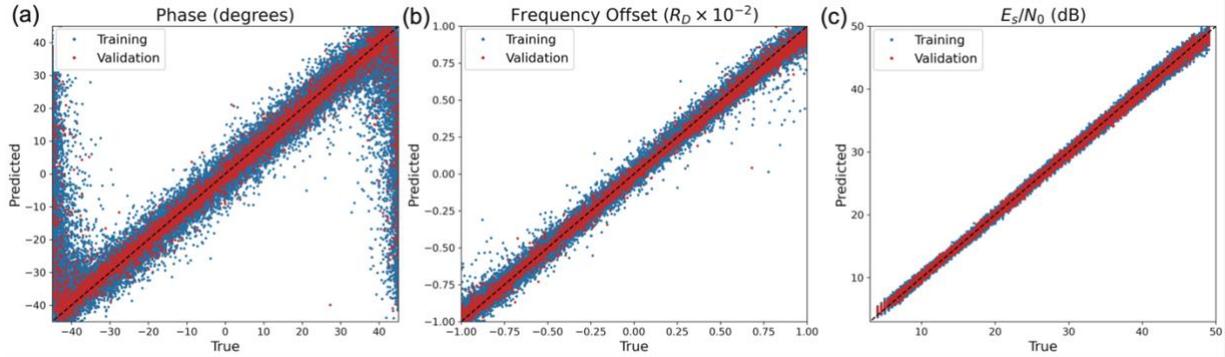

**Fig. 4** Regression of transmission channel parameters using hybrid model. Each training or validation example consists of a series of random bits, which is used to generate a baseband QPSK signal. This signal (Tx) is then put through a physical transmission channel model, resulting in a perturbed version of the original data (Rx). The hybrid model can accurately regress the relative phase offset (a) and the frequency offset, expressed as a percentage of the data rate, $R_D$, (b) between Tx and Rx, as well as the amount of additive white gaussian noise (AWGN), expressed as the energy per symbol to noise power spectral density ratio, $E_s/N_0$, added to the signal from the transmission channel (c).

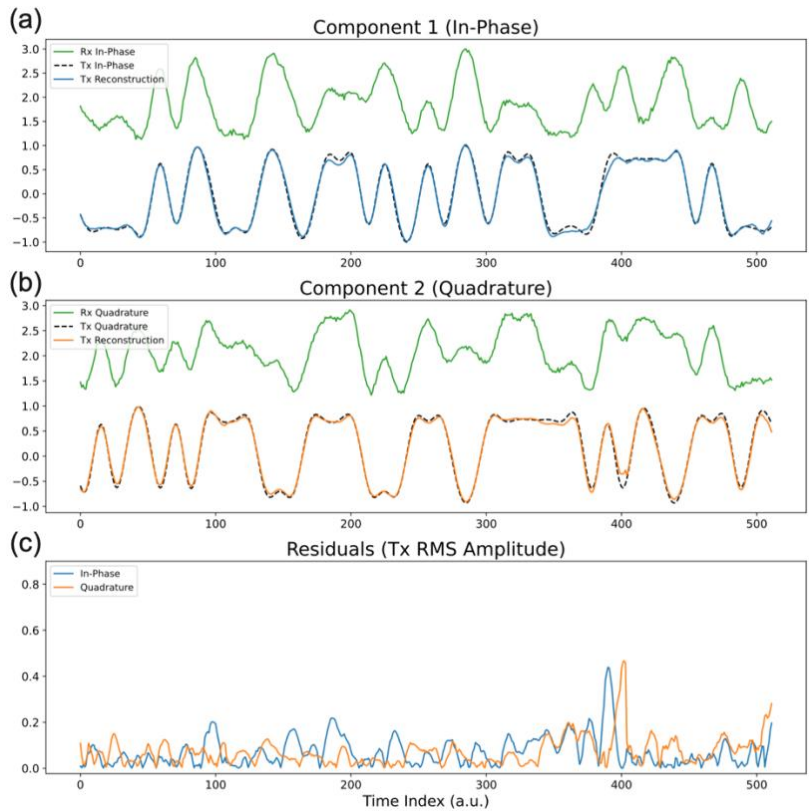

**Fig. 5** Reconstruction of a baseband QPSK signal using decoder layers of the hybrid model. As shown in figure 2, the hybrid model reconstructs the original transmitted signal (Tx) using the received signal (Rx) which has been degraded by the transmission channel. The in-phase (a) and quadrature (b) components for the Rx, Tx, and Tx reconstruction are shown. The residuals shown in (c), defined as the absolute difference between Tx and Tx reconstruction expressed in units of root mean square Tx amplitude, have a mean value of 7 percent.

Fig. 6 shows the performance of a hybrid model trained to classify the total number of symbols in a message. For this task, we generate messages from all 13 modulation types, with a fixed total length (512 samples), and a variable number of symbols (between 16 and 32 symbols per message) and propagate the Tx signal through the harsh transmission channel used in the modulation classification task and discussed in Methods. Figs. 6(a) and 6(b) show the confusion matrices at unnormalized signal-to-noise ratios of 0 and -10 dB, respectively, with a logarithmic color scale. The curve in Fig 6(c) shows the mean accuracy across all classes at each SNR. We note that as the noise level increases and the model's performance begins to degrade, erroneous predictions generally differ from the true value by one symbol, an intuitive result.

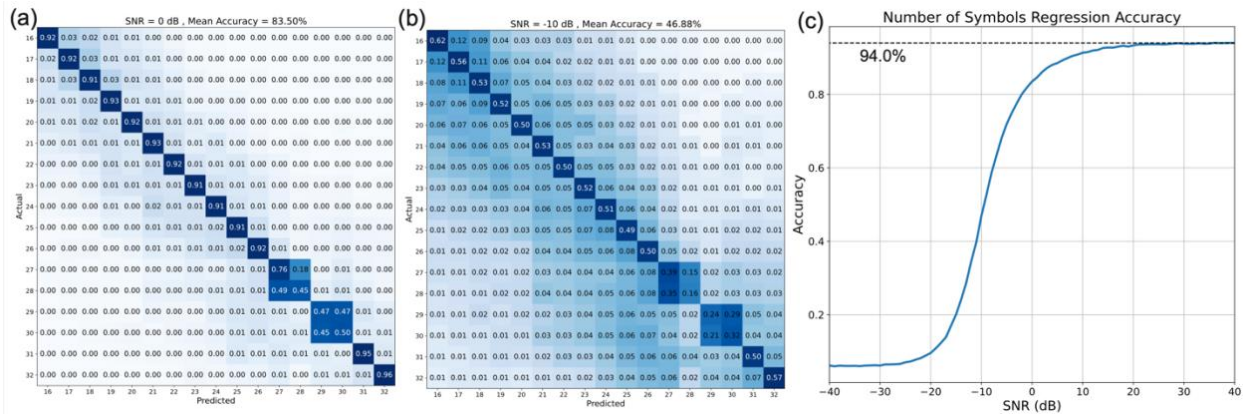

**Fig. 6** Confusion matrices showing performance of the hybrid model at inferring the number of symbols in a message for signal to noise ratios of 0 dB (a) and -10 dB (b), averaged over all 13 modulation types. The curve in (c) shows the models accuracy (averaged over all possible modulation types and symbol numbers) at SNR values between –40 to 40 dB. For high signal to noise ratios, the hybrid model has a mean accuracy of 94.0% when inferring how many symbols are in an average message, as shown in (c).

The final demonstration of the potential for DNNs in the digital signals processing realm is a model trained to demodulate digital messages directly from the Rx baseband data. For this task, we assume a mild channel that approximates a realistic RF communications setup in the presence of light to moderate Rayleigh fading. We assume that the Tx and Rx RF oscillators are tuned to one another (within a tolerance typical of average RF consumer equipment), and that the receiver uses a phase-locked Costas loop to limit the phase shift of the Rx signal. For more details on the mild channel, see Methods.

Fig. 7 shows the hybrid model's accuracy in identifying the symbols of a message when performing direct demodulation of BPSK, QPSK, and 16-QAM modulated input data. The performance of the model is excellent for all modulation schemes with over 99% accuracy for messages with low noise. In practice, messages sent in a Rayleigh channel are usually corrected at the receiver using equalization and are specially designed using forward error correction (FEC) encoding and interleaving to combat the deleterious effects of fading. In contrast, our model is only assuming a phase-locked loop to correct phase and frequency offset with no other channel correcting measures being taken. Integration of FEC codes into a DNN demodulator is a ripe area for future research and could increase model performance to levels that are competitive with commercial communications systems. Nevertheless, our results indicate that direct demodulation of baseband RF signals using DNNs is feasible.

In conclusion, we have presented a powerful new dataset for digital signals processing machine learning tasks in the RF domain. We have also provided a physical transmission channel model which enables the training of models that can accomplish new and novel tasks. We describe a new deep learning model, specially inspired by and designed for the RF domain, which combines an autoencoder convolutional network with a transformer network. We show this hybrid model can efficiently accomplish a variety of RF signal processing tasks, namely, automatic modulation classification, regression of transmission channel parameters, signal denoising and reconstruction, classification of message properties, and direct demodulation of messages.

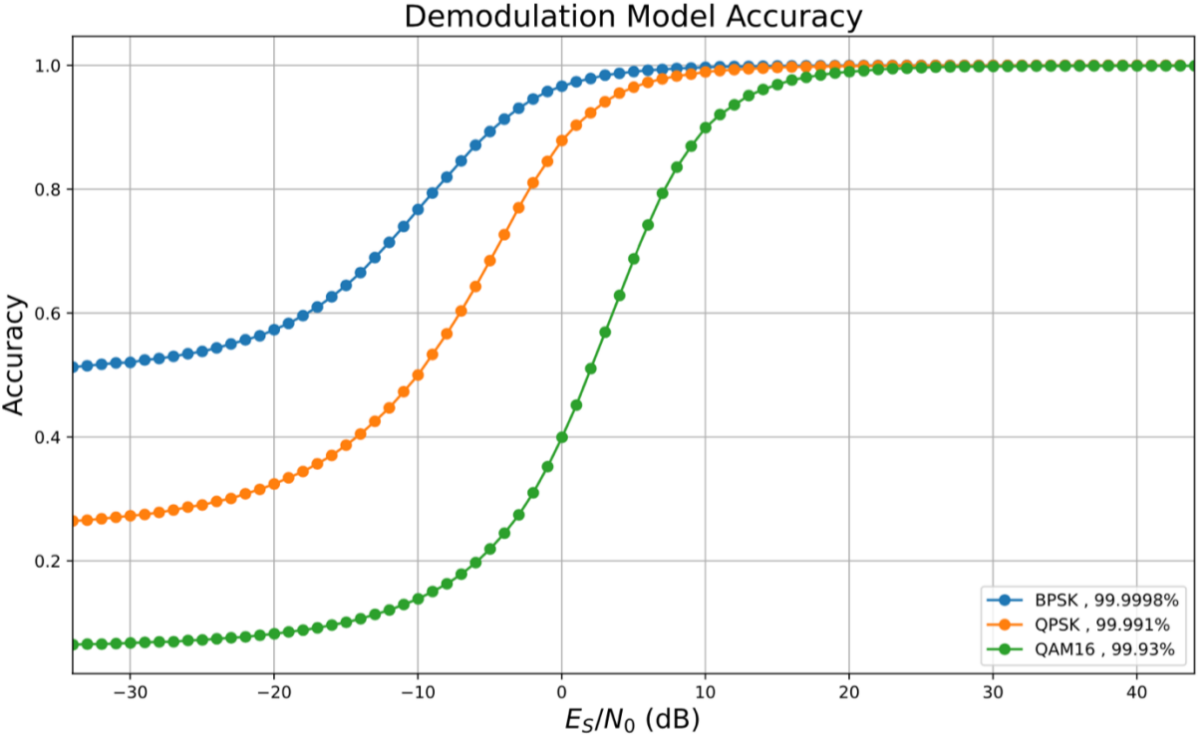

**Fig. 7** Accuracy of hybrid model when demodulating signals for three different modulation types under a channel with Rayleigh fading and additive white gaussian noise (AWGN). The percentages in the legend indicate the model accuracy in the asymptotic limit of high signal-to-noise ratios (expressed here as energy per symbol to noise power spectral density ratio, $E_s/N_0$) for each modulation type.

# Methods

**Data Generation**

All code was written in Python using the PyTorch deep learning framework. Baseband RF data examples are created by the following algorithm: First, an oversampling value (the number of time samples per symbol) is randomly sampled within a predefined range. This oversampling value (along with the predefined shape of the final data vector) is used to determine the number of symbols in the message represented by the data example. Next, a modulation type is selected, and a random, complex-valued message representation is generated from the symbols in that modulation's constellation. Then, an excess bandwidth value for the root-raised-cosine (RRC) filter is randomly sampled (within a predefined range) and the data example is passed through the pulse-shaping, low-pass RRC filter. For all data used in this study, the data rate was used as the base frequency unit.

**Description of Physical Transmission Channel Model**

**Carrier phase and frequency offset:** Consider a transmitter (Tx) and receiver (Rx) that with two independent local RF oscillators. If the two oscillators have a frequency offset of $\Delta f$ and a phase offset of $\varphi$, the in-phase and quadrature components are transformed between Tx and Rx (I(t), Q(t) and I'(t), Q'(t), respectively) according to:

$$\begin{pmatrix} I'(t) \\ Q'(t) \end{pmatrix} = \begin{pmatrix} \cos(2\pi \Delta f t + \varphi) & -\sin(2\pi \Delta f t + \varphi) \\ \sin(2\pi \Delta f t + \varphi) & \cos(2\pi \Delta f t + \varphi) \end{pmatrix} \begin{pmatrix} I(t) \\ Q(t) \end{pmatrix} \quad (1)$$

**Rayleigh Fading:** We use Jakes' model[37] for non-line-of-sight fading to generate a random, complex variable to add to the baseband RF signal. The model assumes that $N$ scatterers are equally spaced at angles $\theta_n = 2\pi n/N$ around the receiver. This complex variable is given by,

$$x + iy = \frac{1}{\sqrt{N}} \sum_{n}^{N} \{\cos(2\pi \eta \tau \cos(\theta_n) + \alpha_n) + i \sin(2\pi \eta \tau \cos(\theta_n) + \beta_n)\} \quad (2)$$

Here N is the number of scatterers to sum over, $\alpha_n$, and $\beta_n$ are random phases, $\eta$ is a dimensionless fading strength parameter, and $\tau$ is the dimensionless time. If the maximum doppler shift is $f_D$, the total elapsed time per message is $T_M$, and the time variable is t, then $\eta = f_D T_M$, and $\tau = t/T_M$. Fading transforms the in-phase and quadrature components between Tx and Rx (I(t), Q(t) and I'(t), Q'(t), respectively) according to:

$$\begin{pmatrix} I'(t) \\ Q'(t) \end{pmatrix} = \begin{pmatrix} x(t) & -y(t) \\ y(t) & x(t) \end{pmatrix} \begin{pmatrix} I(t) \\ Q(t) \end{pmatrix} \quad (3)$$

Throughout the manuscript we refer to three distinct channel regimes: harsh, medium, and mild. We define these channels explicitly here.

**Harsh channel:**

The harsh channel is used for the automatic modulation classification and number of symbols classification tasks. This channel assumes a moderate Rayleigh fading environment, and no channel correction between the transmitter and the ML model input at the receiver. The channel parameters for each training and validation example are drawn from uniform distributions specified in Table 1.

| Name | Symbol | Range (units) |
|---|---|---|
| Fading strength | $\eta$ | [0.1,1.0] |
| Phase Offset | $\varphi$ | [-π,π] (radians) |
| Frequency Offset | $\Delta f$ | [-0.01,0.01] (Data Rate) |
| Signal-to-Noise Ratio | SNR | [-10,30] (dB) |

**Table 1** Summary of parameters for "harsh" transmission channel.

**Medium channel:**

The medium channel is used for the regression of channel parameters task. This channel assumes no Rayleigh fading, and limited channel correction between the transmitter and the ML model input at the receiver which limits the phase offset. The channel parameters for each training and validation example are drawn from uniform distributions specified in Table 2.

| Name | Symbol | Range (units) |
|---|---|---|
| Phase Offset | $\varphi$ | [-π/4,π/4] (radians) |
| Frequency Offset | $\Delta f$ | [-0.01,0.01] (Data Rate) |
| Signal-to-Noise Ratio | SNR | [-2,40] (dB) |

**Table 2** Summary of parameters for "medium" transmission channel.

**Mild channel:**

The mild channel is used for the demodulation task. This channel assumes a moderate Rayleigh fading environment, and realistic channel correction between the transmitter and the ML model input at the receiver (specifically, we assume the presence of a Costas loop). The channel parameters for each training and validation example are drawn from uniform distributions specified in Table 3.

| Name | Symbol | Range (units) |
|---|---|---|
| Fading strength | $\eta$ | [0.1,1.0] |
| Phase Offset | $\varphi$ | [-10,10] (degrees) |
| Frequency Offset | $\Delta f$ | [-1e-4,1e-4] (Data Rate) |
| Signal-to-Noise Ratio | SNR | [-10,40] (dB) |

**Table 3** Summary of parameters for "mild" transmission channel.

## Model Layer Layouts and Definitions

**Encoder Layers:** The encoder consists of alternating Conv1D (a series of Conv1D, BatchNorm1D, and activation layers) and MaxPool1D blocks. Explicit block sizes and parameters are given in Table 4. L refers to the number of time samples in the input vector.

| Block Name | Input | Output | Parameters |
|---|---|---|---|
| Conv1D | (2,L) | (128,L) | kernel size = 13<br>padding = 6<br>activation = ReLU |
| MaxPool1D | (128,L) | (128,L/2) | |
| Conv1D | (128,L/2) | (256,L/2) | kernel size = 13<br>padding = 6<br>activation = ReLU |
| MaxPool1D | (256,L/2) | (256,L/4) | |
| Conv1D | (256,L/4) | (256,L/4) | kernel size = 13<br>padding = 6<br>activation = ReLU |
| Conv1D | (256,L/4) | (256,L/4) | kernel size = 13<br>padding = 6<br>activation = ReLU |

**Table 4** Summary of encoder network layers and parameters.

**Decoder Layers:** The decoder consists of alternating Conv1D (a series of Conv1D, BatchNorm1D, and activation layers) and Upsample blocks. Explicit block sizes and parameters are given in Table 5. L refers to the number of time samples in the input vector.

| Block Name | Input | Output | Parameters |
|---|---|---|---|
| Conv1D | (256,L/4) | (128,L/4) | kernel size = 13<br>padding = 6<br>activation = Leaky ReLU |
| Upsample | (128,L/4) | (128,L/2) | |
| Conv1D | (128,L/2) | (128,L/2) | kernel size = 13<br>padding = 6<br>activation = Leaky ReLU |
| Upsample | (128,L/2) | (128,L) | |
| Conv1D | (128,L) | (64,L) | kernel size = 13<br>padding = 6<br>activation = Leaky ReLU |
| Conv1D | (64,L) | (2,L) | kernel size = 13<br>padding = 6<br>activation = None |

**Table 5** Summary of decoder network layers and parameters.

**Transformer:** We use the Pytorch reformer[35] architecture with a hidden dimension of 256, a depth of 2, and 8 heads. The reformer implementation was chosen because it uses a memory-efficient approximation of the full attention matrix. This approximation results in a model that ultimately uses much less memory with faster performance for long sequences, without a significant reduction in model performance.

**Fully Connected Classifier:** The output of the transformer is flattened and used as input for a fully connected layer with an output size of 1024, followed by BatchNorm1D, ReLU, and Dropout layers. The final layer of the classifier is another fully connected layer whose output size changes depending on the task at hand. The dropout percentage used throughout is 0.2.

**Model Setup and Training**

Consider the model, M, pictured in Fig. 2, which takes as inputs a perturbed signal, Rx, a target signal, Tx, and some labels, L, and returns a reconstructed signal, Tx', and some probabilities, L'. M consists of an encoder network, E, a decoder network, D, and a classifier network, C. The classification loss, $\mathcal{L}_C$, penalizes C for misclassifying a sample's labels (L ≠ L') and the reconstruction loss, $\mathcal{L}_R$, penalizes D for reconstructing a signal that differs from the target signal (Tx ≠ Tx'). A trained model minimizes the total loss,

$$\mathcal{L} = \sum_i \lambda_{C_i} \mathcal{L}_{C_i}(\theta_M) + \lambda_R \mathcal{L}_R(\theta_M) \tag{4}$$

where the $\lambda$ are weights for each individual loss term and $\theta_M$ are parameters of the model. We use the same L2 reconstruction loss function for all tasks,

$$\mathcal{L}_R(\theta_M) = MSE(Tx, Tx'(\theta_M)) = |Tx - Tx'(\theta_M)|^2 \tag{5}$$

Explicit classifier loss functions for each task are given below. Gradients are calculated via back propagation and the total loss is minimized via stochastic gradient descent using the ADAM optimizer with a learning rate of 0.001.

**Automatic modulation classification:** For this task the final output dimension of the classifier is set to 13, the number of distinct modulation class types. The position of the highest probability in the output, L, corresponds to the predicted modulation class. The classification loss has a single term,

$$\sum_i \lambda_{C_i} \mathcal{L}_{C_i}(\theta_M) = \lambda_C CE(L, L'(\theta_M)) \tag{6}$$

where *CE* is the cross-entropy loss function,

$$CE(L, L'(\theta_M)) = -\sum_j^{Nc} v_j(L) \log(L'(\theta_M)) \tag{7}$$

Here, $N_c$ is the number of possible classes (13 in this case) and $v_j(L)$ is the one-hot encoded vector of the true class label.

The model was trained for a total of 128 epochs. For the first 64 epochs, changes to the reconstruction loss were penalized with $\lambda_C=1$ and $\lambda_R=0.001$. For the last 64 epochs, this restraint was removed, and the losses were weighted with $\lambda_C=1$ and $\lambda_R=1$. The data was generated and split using $2^{14}$ and $2^{11}$ examples from each modulation class for training and validation, respectively. All data consisted of 512 time-domain samples with oversampling values randomly varying between 16-32 samples/symbol.

**Regression of channel parameters:** For this task the final output dimension of the classifier is set to 4. In the classification model, the output (L') is a set of probabilities, used to infer a discrete class label. In contrast, the regression model returns an L' whose elements are used to infer a set of continuous parameter values. In this case, the classification loss (Eq. 4), also has 4 terms, one for each of the model outputs

$$\sum_i \lambda_{C_i} \mathcal{L}_{C_i}(\theta_M) = \sum_{i=1}^{4} \lambda_i MSE(L_i, L_i'(\theta_M)) \qquad (8)$$

where *MSE* is the L2 loss function,

$$\mathcal{L}_{C_i}(\theta_M) = MSE(L_i, L_i'(\theta_M)) = |(L_i - L_i'(\theta_M)|^2 \qquad (9)$$

The key to successfully training the regression model is properly constructing L, the targets for the model outputs. We use the following values for the model output targets:

$$L = (\cos(\varphi), \sin(\varphi), 100 * \Delta f, SNR) \qquad (10)$$

These choices scale all values of $L_i$ to have magnitudes close to unity. Also, the decision to split the phase offset, $\varphi$, into two separate trigonometric outputs helps constrain the final value of $\varphi$ to the range $(-\pi, \pi)$ and greatly improves the convergence time and performance of the model.

The model was trained for a total of 500 epochs. The value of each loss term in Eqs. (4) and (8) are monitored throughout the training process and the $\lambda_i$ are updated throughout. A summary of loss weighting is shown in Table 6. Briefly, for the first 150 epochs changes to the reconstruction loss are heavily penalized with no penalty to any other terms. For the next 150 epochs, changes to both reconstruction and SNR are penalized. For the next 100 epochs, the phase and frequency offset terms are slightly adjusted. For the final 100 epochs, all $\lambda_i$ are set to 1. The data was generated and split using $2^{17}$ and $2^{13}$ examples from the QPSK modulation class for training and validation, respectively. All data consisted of 512 time-domain samples with oversampling values randomly varied between 8-16 samples/symbol.

| Epoch Numbers | $\lambda_R$ | $\lambda_{C1}(\cos(\varphi))$ | $\lambda_{C2}(\sin(\varphi))$ | $\lambda_{C3}(100*\Delta f)$ | $\lambda_{C4}(SNR)$ |
|---|---|---|---|---|---|
| 0-150 | 0.001 | 1 | 1 | 1 | 1 |
| 150-300 | 0.001 | 1 | 1 | 1 | 0.01 |
| 300-400 | 0.001 | 1 | 1 | 0.2 | 0.01 |
| 400-500 | 1 | 1 | 1 | 1 | 1 |

**Table 6** Summary of loss term weighting during training epochs for regression of channel parameters task.

**Number of message symbols classification:** The model was trained for 8 epochs with $\lambda_C = \lambda_R = 1$ for all epochs. The data was generated and split using $2^{14}$ and $2^{11}$ examples from each of the 13 modulation classes for training and validation, respectively. All data consisted of 512 time-domain samples with the number of symbols per message being drawn from a uniform distribution and varying between 16-32 symbols/message. For this task the oversampling value is derived from the number of symbols per message instead of being directly drawn from a uniform distribution, as is the case for all other tasks.

The final output dimension of the classifier is set to 17, the number of possible symbols in a message. The position of the highest probability in the output, L, corresponds to the predicted number of symbols/message. The classification loss has a single term,

$$\sum_i \lambda_{C_i} \mathcal{L}_{C_i}(\theta_M) = \lambda_C CE(L, L'(\theta_M)) \quad (11)$$

where *CE* is the cross-entropy loss function,

$$CE(L, L'(\theta_M)) = -\sum_j^{N_c} v_j(L) \log(L'(\theta_M)) \quad (12)$$

Here, $N_c$ is the number of possible classes (17 in this case) and $v_j(L)$ is the one-hot encoded vector of the true class label.

**Signal demodulation:**

For this task the final output dimension of the classifier is set to 256×k, where k is the number of symbols in the modulation constellation. Each data example consists of 256 symbols and each symbol is treated as an individual k-class classification problem. In this case, the classification loss (Eq. 4), has 256 terms, one for each symbol of the message

$$\sum_i \lambda_{C_i} \mathcal{L}_{C_i}(\theta_M) = \lambda_C \sum_{i=1}^{256} CE(L_i, L_i'(\theta_M)) \quad (13)$$

where *CE* is the cross-entropy loss function,

$$CE(L, L'(\theta_M)) = -\sum_j^k v_j(L) \log(L'(\theta_M)) \quad (14)$$

Here, k is the number of possible classes (2 for BPSK, 4 for QPSK, and 16 for 16-QAM) and $v_j(L)$ is the one-hot encoded vector of the true message symbols.

The model was trained for a total of 256 epochs with $\lambda_C = 1$ and $\lambda_R = 0.01$ for the first 128 epochs and $\lambda_C = \lambda_R = 1$ for the final 128 epochs. The data was generated and split using $2^{16}$ and $2^{13}$ examples for training and validation, respectively for each modulation type. All data consisted of 1024 time-domain samples with an oversampling of 4.

**Data Availability**

Code for dataset generation and transmission channel modeling are available at https://github.com/pnnl/DieselWolf. Model definitions and analysis code that were used in this study are available from the corresponding authors upon reasonable request.

**Acknowledgements**

This research was supported by the Pacific Northwest National Laboratory (PNNL) Laboratory Directed Research and Development program. PNNL is operated for DOE by Battelle Memorial Institute under contract DE-AC05-76RL01830.

**Author contributions**

B.S. and M.G. conceived the project. B.S. created the dataset and transmission channel model. Y.W. developed the hybrid model. B.S., Y.W., N.M., and M.G. trained and tested models, performed statistical analysis, and generated figures. B.S. and Y.W. wrote the manuscript. M.G. supervised the project. All authors reviewed the manuscript.

**Competing Interests**

The authors declare no competing interests.